\documentclass[runningheads]{llncs}

\usepackage[T1]{fontenc}
\usepackage{amsmath, amsfonts, amssymb}
\usepackage{tikz} \usetikzlibrary{calc, positioning, arrows.meta}
\usepackage{ring3dir}
\usepackage{ring5ring}
\usepackage{fibdigraph}
\usepackage{stackring}
\usepackage{subcaption}
\usepackage{graphicx} \graphicspath{ {./figures/} }
\usepackage{hyperref}
\usepackage{xcolor}  
\usepackage{cleveref}
\usepackage{nicefrac}
\usepackage{algorithm}
\usepackage{algorithmic}
\usepackage{bm}
\usepackage{microtype}
\usepackage{orcidlink} \renewcommand{\orcidID}[1]{\orcidlink{#1}}
\usepackage{mathtools}
\usepackage{enumerate}
\usepackage{pgfplots}
\usepackage{multirow}

\usetikzlibrary{calc,angles,quotes}

\renewcommand{\phi}{\varphi}
\newcommand{\tsp}[1]{\operatorname{TSP}(#1)}
\newcommand{\dfj}[1]{\operatorname{DFJ}(#1)}
\newcommand{\R}{{R}}
\newcommand{\bb}[1]{\operatorname{BB}(#1)}
\newcommand{\fs}[1]{\operatorname{FS}(#1)}
\newcommand{\sall}{\mathcal{S}_\mathrm{all}}

\newcommand{\sep}[1]{\operatorname{SEP}(#1)}

\newcommand{\suchthat}{\;\ifnum\currentgrouptype=16 \middle\fi|\;}

\renewcommand{\setminus}{\!\smallsetminus\!}

\title{TSP instances that require exponentially\\ many subtour elimination constraints}
\titlerunning{TSP instances that require exponentially many SECs}

\author{Eleonora Vercesi\inst{1,2}\orcidID{0000-0002-1621-2484} \and Austin Buchanan\inst{3}\orcidID{0000-0003-2999-9666}}

\institute{Università della Svizzera Italiana (USI)
\and Dalle Molle Institute for Artificial Intelligence USI-SUPSI
\and Oklahoma State University}

\begin{document}
\maketitle

\begin{abstract}
We construct pathological TSP instances that require exponentially many subtour elimination constraints (SECs) in the sense that, if a smaller collection of SECs is imposed, then the Dantzig-Fulkerson-Johnson model returns a disconnected solution costing less than an optimal tour. For asymmetric instances on $n$ vertices, we prove a growth rate of $\Omega(\phi^n)$, where $\phi\approx 1.618$ is the golden ratio. A node-splitting reduction gives metric instances with $\Omega(\phi^{n/2})$ growth. We also give instances in the plane under the Euclidean, Rectilinear, and Maximum metrics that require exponentially many SECs. Concorde struggles on them, with solve times growing exponentially despite tiny integrality gaps.
\end{abstract}

\section{Introduction}

The classic TSP formulation of Dantzig, Fulkerson, and Johnson (DFJ)~\cite{dantzig1954} can be written as follows, where $G=(V,E)$ is a complete undirected graph on $n$ vertices and $c_e$ is the cost to traverse edge $e\in E$. Traditionally, at least on paper, the model imposes exponentially many subtour elimination constraints (SECs)~\eqref{sec}, one for each set $S$ in the set family $\sall\coloneqq \{ S \subset V \suchthat 2 \le |S| \le n-2\}$. However, for notational convenience, we parameterize the DFJ model by $\mathcal{S} \subseteq \sall$.
\begin{subequations}
\begin{align}
\dfj{G,\mathcal{S}}=\min & \sum_{e \in E} c_e x_e \\
\text{s.t.}&\sum_{e \in \delta(v)} x_e = 2 & \forall v \in V\label{deg2}\\
& \sum_{e \in \delta(S)} x_e \ge 2 & \forall S \in \mathcal{S} \label{sec}\\
& x_e \in \{0,1\} & \forall e \in E. \label{eq:dfj_int}
\end{align}\label{dfj}
\end{subequations}
Here, $\delta(S)$ denotes the subset of edges with one endpoint in $S$. In a slight abuse of notation, for singleton sets $S=\{v\}$, we simply write $\delta(v)$ rather than $\delta(\{v\})$. 

In practice, a tiny fraction of the exponentially many SECs often suffices to prove optimality~\cite{miliotis1976,pferschy2017}. Indeed, Vercesi and Buchanan~\cite{vercesi2026} show that the classic \texttt{dantzig42} instance of Dantzig, Fulkerson, and Johnson requires only four SECs. That is, there exists a set family $\mathcal{S}=\{S_1, S_2, S_3, S_4\}$ 
for which the DFJ model over just those four SECs (solved as an integer program) has objective value equal to the length of an optimal tour, denoted $\tsp{G}$. Moreover, there is no smaller set family with this property. These notions are formalized in Definition~\ref{suff}. 

\begin{definition}\label{suff}
A set family $\mathcal{S}\subseteq \sall$ is {sufficient} if $\dfj{G,\mathcal{S}}=\tsp{G}$, and the number of required SECs is the size of a smallest sufficient set family, i.e.,
\begin{align*}
\R(G)\coloneqq\min~ & | \mathcal{S}| \\
\mathrm{s.t.}~ & \dfj{G,\mathcal{S}} = \tsp{G} \\
 & \mathcal{S} \subseteq \sall.
\end{align*}
\end{definition}

By this definition, an instance $G$ \emph{requires} $\R(G)$ subtour constraints in the sense that all set families $\mathcal{S}$ of cardinality less than $\R(G)$ yield cost strictly less than TSP optimal. Vercesi and Buchanan~\cite{vercesi2026} computed $\R(G)$ for TSPLIB instances, finding that it is often small. Indeed, across all TSPLIB instances with fewer than 100 vertices, 64\% of them have $\R(G)\le 5$, and 91\% have $\R(G)\le 12$. Meanwhile, HardTSPLIB instances~\cite{vercesi2023generation}, which are hard for the Concorde TSP solver, require significantly more, with a particular 30-vertex instance requiring 235 SECs. Similarly, some of the most challenging metric TSP instances from the literature, the Rectilinear 3-dimensional instances of Zhong~\cite{zhong2025}, empirically seem to require a quadratic number of SECs, $\Theta(n^2)$. Still, it remained an open question whether any class of instances requires {exponentially many} SECs, $2^{\Omega(n)}$. 

{\bf {Our contributions.}} In this paper, we answer this question in the affirmative: there \emph{do} exist pathological TSP instances that require exponentially many SECs. Section~\ref{sec:simplest} gives our simplest construction, the \emph{Ball Bearing} construction, which requires $2^{n/5}$ SECs and can be embedded in the plane, yielding TSP instances under the Euclidean, Rectilinear, and Maximum metrics that also require $2^{n/5}$ SECs. Section~\ref{sec:fib} gives a more complex \emph{Fibonacci} construction that increases the exponential growth to $\Omega(\phi^n)$ for asymmetric TSP and $\Omega(\phi^{n/2})$ for metric TSP, where $\phi\approx 1.618$ is the golden ratio. 
Section~\ref{sec:experiments} applies the Concorde solver~\cite{applegate2003,applegate2007} to the constructions. Empirically, solve times grow exponentially despite tiny integrality gaps. 
This stands in contrast with HardTSPLIB and Rectilinear 3-dimensional instances, which stymie Concorde with their large integrality gaps. 


\section{Ball Bearing Construction}\label{sec:simplest}

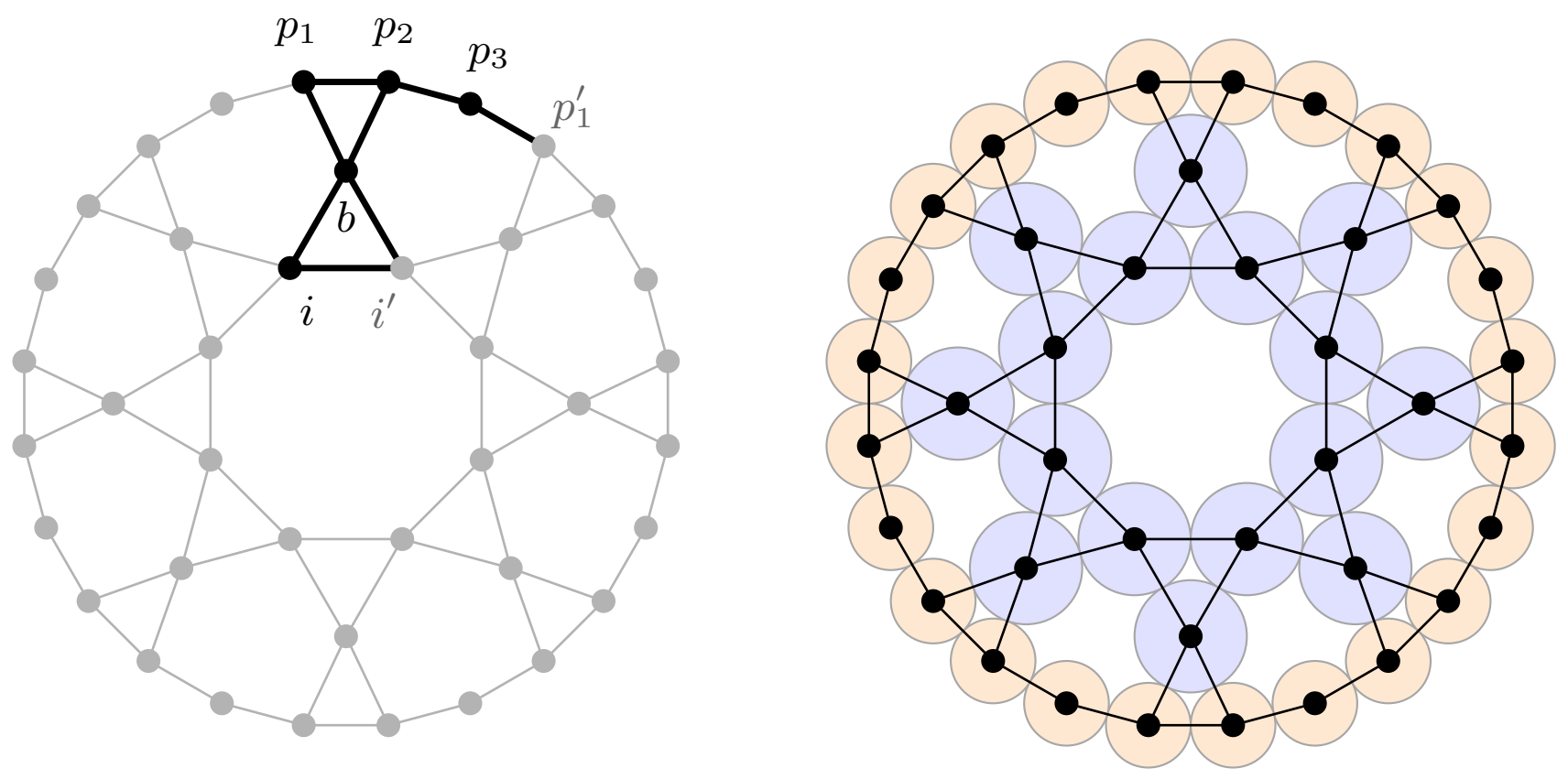
\begin{figure}[ht]
  \begin{subfigure}{0.5\linewidth}
    \centering
    \begin{tikzpicture}
      \TwoNFiveRing[gadgets=8, coins, show coins=false, curved=false, highlight=0, light=black!30, labels=false, width=\linewidth, labels=true, inner label=i, pearl label=b, next inner label={i'}]
    \end{tikzpicture}
    \caption{The graph $\bb{8}$.}
    \label{bba}
  \end{subfigure}\hfill
  \begin{subfigure}{0.5\linewidth}
    \begin{tikzpicture}
      \TwoNFiveRing[gadgets=8, coins, curved=false, highlight=-1, light=black, labels=false, width=\linewidth]
    \end{tikzpicture}
    \caption{A coin embedding for Euclidean metric.}
    \label{bbb}
  \end{subfigure}
  \caption{The Ball Bearing construction.}
  \label{fig:ballbearing}
\end{figure}

This section constructs pathological TSP instances from Ball Bearing graphs, denoted $\bb{k}$. As depicted in Fig.~\ref{bba}, they are obtained by arranging $k\ge 3$ gadgets in a ring. Each gadget has five vertices, with an inner vertex $i$, ball vertex $b$, and three outer (or peripheral) vertices $p_1, p_2, p_3$, yielding $n=5k$ total vertices. Within a gadget $g=\{i,b,p_1,p_2,p_3\}$, there are five edges $\{ ib, bp_1, bp_2, p_1p_2, p_2p_3\}$, and the gadget is connected to the next gadget $g'=\{i',b',p'_1,p'_2,p'_3\}$ in the ring by three edges $\{ii', bi', p_3p'_1\}$. In total, $\bb{k}$ has $8k$ edges.

The following lemma refers to 2-factors. In a graph $G=(V,E)$, an edge subset $F\subseteq E $ is called a 2-factor if $|F \cap \delta(v)|=2$ for all $v \in V$. 


\begin{lemma}\label{notone}
In every 2-factor of $\bb{k}$, every ball vertex $b$ neighbors both vertices from $\{i,i'\}$ or both vertices from $\{p_1,p_2\}$; it cannot neighbor one from each.
\end{lemma}

\begin{proof}
Since every peripheral vertex (of type) $p_3$ has degree two in $\bb{k}$, its two incident edges belong to every 2-factor $F$. Accordingly, one edge is forced incident to $p_1$ (by prior gadget) and one to $p_2$. If the edge $p_1p_2$ between them belongs to $F$, then $p_1$ and $p_2$ already have degree two, forcing $b$ to pick $bi$ and $bi'$. Otherwise, if edge $p_1p_2$ does \emph{not} belong to $F$, then $p_1$ and $p_2$ each have one remaining edge to select ($bp_1$ and $bp_2$), so $b$ neighbors both $p_1$ and $p_2$ in $F$. \qed
\end{proof}

\begin{lemma}\label{ham}
The graphs $\bb{k}$ are themselves non-Hamiltonian but admit $2^{n/5}$ vertex bipartitions $\{S, \bar S\}$ for which $S$ and $\bar S$ induce Hamiltonian subgraphs.
\end{lemma}

\begin{proof}
Let $F$ be a 2-factor of $\bb{k}$. Consider an inner vertex $i$ and peripheral vertex $p_3$. By Lemma~\ref{notone}, the cycle $C\subseteq F$ that contains $p_3$ cannot cross into the inner ring, so $i$ belongs to a different cycle, thus $F$ is not a Hamiltonian cycle.

Denote by $B$ the set of ball vertices and $I$ the inner vertices. Every vertex subset $S$ with $I \subseteq S \subseteq I \cup B$ and its complement $\bar S = V(\bb{k})\setminus S$ induce a Hamiltonian subgraph (by snaking along the inner ring and outer ring, respectively, picking up ball vertices from the given side). There are $2^{|B|}=2^{n/5}$ such sets. \qed
\end{proof}

By giving $\bb{k}$'s edges weight one, and missing edges weight two, we arrive at a 1,2-TSP instance that requires $2^{k}=2^{n/5}$ SECs. To prove this, we recall a general notion (Definition~\ref{cheapdef}) and result (Proposition~\ref{cheapbound}) that we use throughout. 

\begin{definition}\label{cheapdef}
A subset $S\subset V$ is cheap if $\tsp{G[S]}+\tsp{G[\bar S]} < \tsp{G}$.
\end{definition}

If a vertex subset $S\subset V$ is cheap, then every sufficient set family $\mathcal{S}$ contains either $S$ or its complement $\bar S$ (and the associated SECs are equivalent under the degree-two constraints~\eqref{deg2}). This follows, for example, from Vercesi and Buchanan~\cite{vercesi2026}'s characterization of sufficient set families (see their Theorem 4 and their \S4.2.1). 
The idea here is that, if $\mathcal{S}$ contains neither $S$ nor its complement $\bar S$, then the union of a TSP tour of $G[S]$ and a TSP tour over $G[\bar S]$ will satisfy the constraints of $\dfj{G,\mathcal{S}}$, as these are the only two constraints from the full DFJ model that would ``cut off'' such a solution. (This statement applies to the symmetric version of the DFJ model given in the introduction, as well as to its asymmetric variant.) This implies the following proposition.

\begin{proposition}\label{cheapbound}
If $\mathcal{C}$ is a collection of cheap sets of $G$, no two of which partition $V$, then $G$ requires at least $|\mathcal{C}|$ SECs. 
\end{proposition}

\begin{theorem}\label{bb12}
The 1,2-TSP instance derived from $\bb{k}$ requires at least $2^{n/5}$ SECs.
\end{theorem}

\begin{proof}
By Lemma~\ref{ham}, $\bb{k}$ is not Hamiltonian, so the associated 1,2-TSP instance $G$ has cost $\tsp{G}\ge n+1$. Meanwhile, also by Lemma~\ref{ham}, $\bb{k}$ admits $2^{n/5}$ vertex bipartitions $\{S, \bar S\}$ for which $S$ and $\bar S$ induce Hamiltonian subgraphs, so the 1,2-weighting has $\tsp{G[S]}+\tsp{G[\bar S]}=n$. Since $G$ admits $2^{n/5}$ cheap sets, the result follows by Proposition~\ref{cheapbound}. \qed
\end{proof}

Denote by $\sep{G}$ the optimal value of the linear program (LP)~\eqref{dfj} in which the integrality constraint~\eqref{eq:dfj_int} is replaced by $x_e \in [0, 1]$ and all SECs are imposed (i.e., $\mathcal{S}=\mathcal{S}_\textrm{all}$). This LP is the so-called \textbf{S}ubtour \textbf{E}limination \textbf{P}roblem.
%
%
The \emph{integrality gap} of a TSP instance is defined as the ratio between $\tsp{G}$ and $\sep{G}$; this number is always between $1$ and $\frac{3}{2} - 10^{-36}$ for metric instances~\cite{karlin2022,karlin2021,karlin2023}. 
In the literature, computationally hard-to-solve instances are associated with large integrality gaps~\cite{hougardy2021hard,vercesi2023generation,zhong2025}. 
However, the 1,2-TSP instances derived from $\bb{k}$ have tiny integrality gaps, as noted in Proposition~\ref{prop:bb_ig_sep} and proved in Appendix~\ref{appendix:sep}.

\begin{proposition}\label{prop:bb_ig_sep}
The 1,2-TSP instances $G$ derived from $\bb{k}$ have
\begin{align*}
\frac{\tsp{G}}{\sep{G}} = \frac{n + 1}{n} \to 1~\text{as}~n \to \infty.
\end{align*}
\end{proposition}

\subsection{Planar Embeddings}

The Ball Bearing graphs are planar, and it is known by the Koebe-Andreev-Thurston theorem~\cite[Ch.5]{lovasz2019} that every planar graph admits a \emph{coin embedding}, where vertices are represented by non-overlapping circular disks in the plane, with two disks being tangent precisely when their vertices are connected by an edge. Fig.~\ref{bbb} gives an example. This immediately gives the following corollary.

\begin{corollary}\label{corcoin}
The Euclidean TSP instance derived from a coin embedding of $\bb{k}$ requires at least $2^{n/5}$ SECs.
\end{corollary}

\begin{proof}
Let $r_v$ be the radius of vertex $v$ in the coin embedding. Then, the 2-factors from Lemma~\ref{ham} all have cost $f\coloneqq\sum_{v \in V} 2r_v$, since they enter and exit each coin once, never traversing a gap. These $2^{n/5}$ sets are cheap since any TSP tour must use at least one edge from outside $\bb{k}$ (due to non-Hamiltonicity, Lemma~\ref{ham}), forcing an optimal tour to traverse a gap between non-touching coins, while \emph{still} having to traverse each coin's radius twice, forcing $\tsp{G} > f$. \qed
\end{proof}

The particular coin embedding depicted in Fig.~\ref{bbb} ``works'' for all $k\ge 3$, where the inner and ball coins can be taken as unit disks. Their positions, as well as the radii of the outer coins, follow from this choice. For the details, refer to the accompanying code, which includes generators for all proposed embeddings. 

When moving to the Maximum and Rectilinear metrics, we would like a general-purpose hammer like the Koebe-Andreev-Thurston theorem saying that all planar graphs admit embeddings as non-overlapping square coins or diamond coins, respectively. Such a theorem does not exist. Nevertheless, we construct suitable embeddings for the Ball Bearing graphs. 

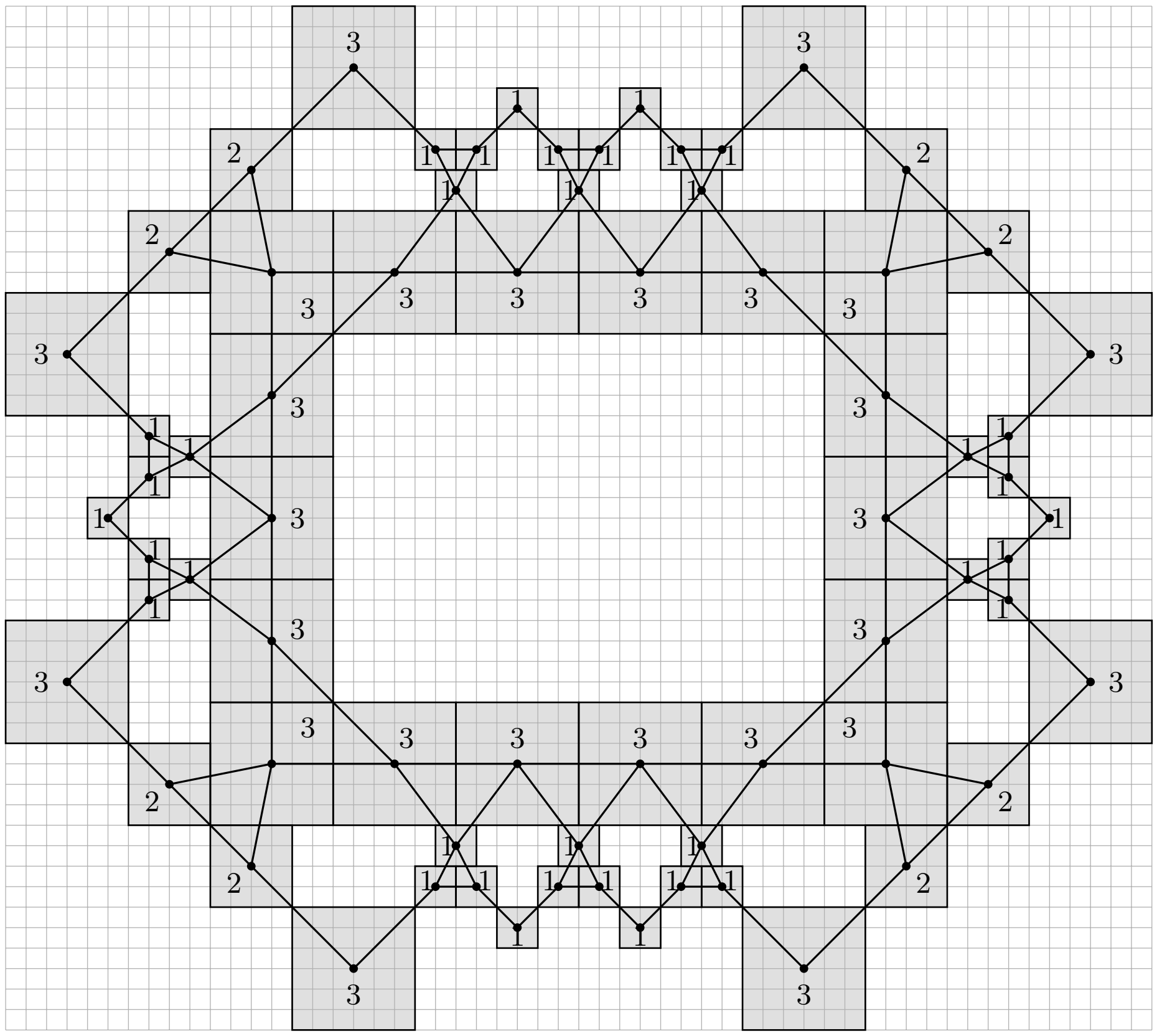
\begin{figure}[ht]
\centering 
\begin{tikzpicture}
\StackRing[a=3, b=2, small, grid, width=\linewidth]
\end{tikzpicture}
\caption{A square coin embedding of $\bb{14}$ for Maximum metric.}
\label{fig:simplestmax}
\end{figure}


%
Fig.~\ref{fig:simplestmax} shows an embedding of $\bb{14}$ as square coins for the Maximum metric. The numbers in the figure indicate the radii, and each square is centered at integer-valued $(x,y)$-coordinates. So, the distances between any two vertices $(x_1,y_1)$ and $(x_2,y_2)$ under the Maximum metric, i.e., $\max \{ |x_2-x_1|, |y_2-y_1|\}$, will also be integer-valued. Although the figure only shows the case of $k=14$ gadgets, the same idea applies for any even $k \ge 8$ (by adding or subtracting pairs of gadgets on the straightaways). Fig.~\ref{fig:simplestmax8} of Appendix~\ref{appendix:figs} depicts the $k=8$ case. 

By rotating the construction 45 degrees, we obtain a suitable embedding for the Rectilinear metric, see Fig.~\ref{fig:simplestrect} of Appendix~\ref{appendix:figs} for $\bb{10}$. Algebraically, this means mapping the $(x,y)$-coordinates of each square to $(\frac{x+y}{2},\frac{x-y}{2})$ for the corresponding diamond. These coordinates will be half-integral. If desired, multiply by two to achieve integer-valued coordinates and distances under the Rectilinear metric, i.e., $|x_2-x_1|+|y_2-y_1|$. Theorem~\ref{thm:maxrect} summarizes these observations.

\begin{theorem}\label{thm:maxrect}
For every even $k\ge 8$, there exists a TSP instance in the plane on $n=5k$ vertices, under the Maximum and Rectilinear metrics, that requires at least $2^{n/5}$ SECs. Moreover, the $(x,y)$-coordinates of each vertex are small integers (specifically, linear in $n$); the distances between vertices are too. 
\end{theorem}

\begin{proof}
Analogous to the proof of Corollary~\ref{corcoin}. \qed
\end{proof}

\section{Fibonacci Construction}\label{sec:fib}

The previous section gave a relatively simple construction based on the Ball Bearing graphs, which allowed us to easily obtain TSP instances in the plane under the Euclidean, Rectilinear, and Maximum metrics. 

Here, we give a more complex construction. We start with a \emph{directed} graph, the {Fibonacci-Sum digraph} $\fs{k}$, which, like before, is non-Hamiltonian, but admits exponentially many vertex bipartitions $\{S, \bar S\}$ that induce Hamiltonian subgraphs. By 1,2-weighting, this gives an ATSP instance that requires on the order of $\phi^n\approx 1.618^n$ SECs. Then, by applying a node-splitting reduction, similar to that of Jonker and Volgenant~\cite{jonker1983}, which doubles the number of nodes, we obtain a symmetric (indeed metric) instance that requires on the order of $\phi^{n/2}$ SECs. This gives substantially faster growth in SECs, $\phi^n\approx 2^{0.6942 n}$ and $\phi^{n/2}\approx 2^{0.3471n}$, than the Ball Bearing construction, $2^{n/5}=2^{0.2n}$.

\begin{figure}[ht]
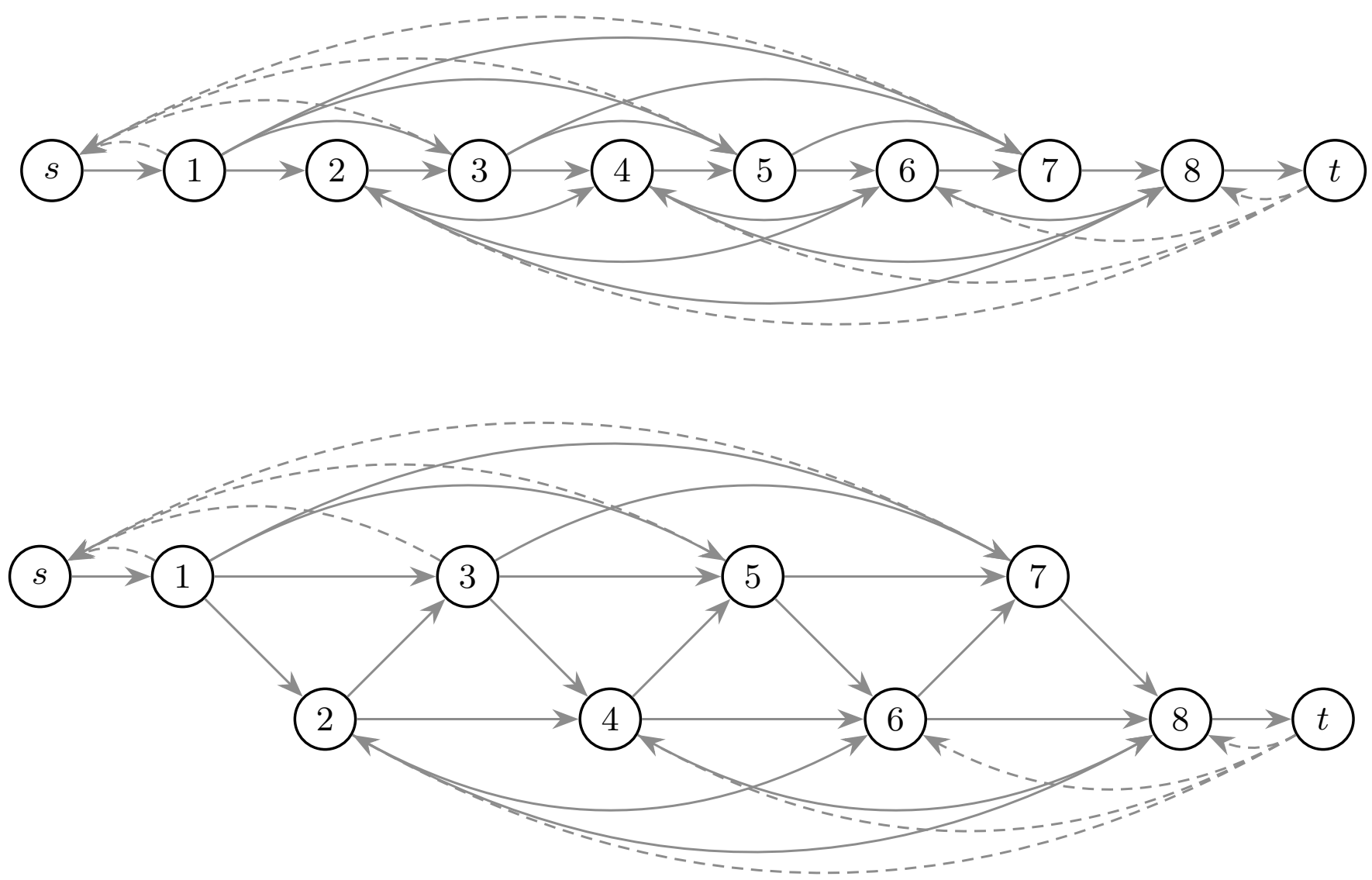

\centering
\resizebox{0.95\columnwidth}{!}
{
\fibgraph[n=10, ends=row, layout=line]} \\
\resizebox{0.95\columnwidth}{!}
{\fibgraph[k=8]
}
\caption{Two drawings of the Fibonacci-Sum digraph $\fs{8}$.}
\label{fig:fibgraph}
\end{figure}

As depicted in Fig.~\ref{fig:fibgraph}, the Fibonacci-Sum digraph $\fs{k}$ is obtained by first arranging an even number $n=k+2$ of vertices in a directed path $(s,1,2,\dots, k,t)$. For every $i,j\in [k]$ with $i < j$, we add edge $(i,j)$ if $i$ and $j$ have the same parity (i.e., odd or even). So far, this gives a directed acyclic graph with $k+1$ \emph{path} edges and $2\binom{k/2}{2}$ \emph{parity} edges. Last, we add $k$ \emph{back} edges, shown in dashes, which connect every odd $i \in [k]$ back to $s$, and $t$ back to every even $j \in [k]$. The second drawing in Fig.~\ref{fig:fibgraph} emphasizes the digraph's odd/even nature.

\tikzset{fib back/.style={densely dashed}}
\fibsetup{hub dx=0pt,  hub dy=0pt}      

Before giving our main theorem, we review two key facts about Fibonacci numbers. Let $F_k$ be the $k$-th Fibonacci number, with $F_1=F_2=1$ and $F_j=F_{j-1}+F_{j-2}$ for $j\ge 3$. It is known that $F_k \ge \phi^{k-2}$, see Cormen et al.~\cite[Lemma 19.3]{cormen2009}. An interesting fact about the Fibonacci numbers is that $F_k$ equals the number of \emph{compositions} of $k$ into odd parts, see Stanley~\cite[Exercise 1.35(d)]{stanley2011enumerative}. For instance, Table~\ref{tab:k8} gives the $F_8=21$ compositions of $8$ into odd parts.

\begin{table}[ht]
\centering
\small
\caption{The $F_8=21$ compositions of $k=8$ into odd parts.}
\label{tab:k8}
\begin{tabular}{l|l|l|l|l}
\# ~& composition & runs & $S$ & $\bar S$\\ \hline
1 & $1+7$ & {1}\,$|$\,{2345678} & $\{s,1\}$ & $\{2,3,4,5,6,7,8,t\}$ \\
2 & $3+5$ & {123}\,$|$\,{45678} & $\{s,1,2,3\}$ & $\{4,5,6,7,8,t\}$ \\
3 & $5+3$ & {12345}\,$|$\,{678} & $\{s,1,2,3,4,5\}$ & $\{6,7,8,t\}$ \\
4 & $7+1$ & {1234567}\,$|$\,{8} & $\{s,1,2,3,4,5,6,7\}$ & $\{8,t\}$ \\
5 & $1+1+1+5$ & {1}\,$|$\,{2}\,$|$\,{3}\,$|$\,{45678} & $\{s,1,3\}$ & $\{2,4,5,6,7,8,t\}$ \\
6 & $1+1+3+3$ & {1}\,$|$\,{2}\,$|$\,{345}\,$|$\,{678} & $\{s,1,3,4,5\}$ & $\{2,6,7,8,t\}$ \\
7 & $1+1+5+1$ & {1}\,$|$\,{2}\,$|$\,{34567}\,$|$\,{8} & $\{s,1,3,4,5,6,7\}$ & $\{2,8,t\}$ \\
8 & $1+3+1+3$ & {1}\,$|$\,{234}\,$|$\,{5}\,$|$\,{678} & $\{s,1,5\}$ & $\{2,3,4,6,7,8,t\}$ \\
9 & $1+3+3+1$ & {1}\,$|$\,{234}\,$|$\,{567}\,$|$\,{8} & $\{s,1,5,6,7\}$ & $\{2,3,4,8,t\}$ \\
10 & $1+5+1+1$ & {1}\,$|$\,{23456}\,$|$\,{7}\,$|$\,{8} & $\{s,1,7\}$ & $\{2,3,4,5,6,8,t\}$ \\
11 & $3+1+1+3$ & {123}\,$|$\,{4}\,$|$\,{5}\,$|$\,{678} & $\{s,1,2,3,5\}$ & $\{4,6,7,8,t\}$ \\
12 & $3+1+3+1$ & {123}\,$|$\,{4}\,$|$\,{567}\,$|$\,{8} & $\{s,1,2,3,5,6,7\}$ & $\{4,8,t\}$ \\
13 & $3+3+1+1$ & {123}\,$|$\,{456}\,$|$\,{7}\,$|$\,{8} & $\{s,1,2,3,7\}$ & $\{4,5,6,8,t\}$ \\
14 & $5+1+1+1$ & {12345}\,$|$\,{6}\,$|$\,{7}\,$|$\,{8} & $\{s,1,2,3,4,5,7\}$ & $\{6,8,t\}$ \\
15 & $1+1+1+1+1+3$ & {1}\,$|$\,{2}\,$|$\,{3}\,$|$\,{4}\,$|$\,{5}\,$|$\,{678} & $\{s,1,3,5\}$ & $\{2,4,6,7,8,t\}$ \\
16 & $1+1+1+1+3+1$ & {1}\,$|$\,{2}\,$|$\,{3}\,$|$\,{4}\,$|$\,{567}\,$|$\,{8} & $\{s,1,3,5,6,7\}$ & $\{2,4,8,t\}$ \\
17 & $1+1+1+3+1+1$ & {1}\,$|$\,{2}\,$|$\,{3}\,$|$\,{456}\,$|$\,{7}\,$|$\,{8} & $\{s,1,3,7\}$ & $\{2,4,5,6,8,t\}$ \\
18 & $1+1+3+1+1+1$ & {1}\,$|$\,{2}\,$|$\,{345}\,$|$\,{6}\,$|$\,{7}\,$|$\,{8} & $\{s,1,3,4,5,7\}$ & $\{2,6,8,t\}$ \\
19 & $1+3+1+1+1+1$ & {1}\,$|$\,{234}\,$|$\,{5}\,$|$\,{6}\,$|$\,{7}\,$|$\,{8} & $\{s,1,5,7\}$ & $\{2,3,4,6,8,t\}$ \\
20 & $3+1+1+1+1+1$ & {123}\,$|$\,{4}\,$|$\,{5}\,$|$\,{6}\,$|$\,{7}\,$|$\,{8} & $\{s,1,2,3,5,7\}$ & $\{4,6,8,t\}$ \\
21 & $1+1+1+1+1+1+1+1$ & {1}\,$|$\,{2}\,$|$\,{3}\,$|$\,{4}\,$|$\,{5}\,$|$\,{6}\,$|$\,{7}\,$|$\,{8} & $\{s,1,3,5,7\}$ & $\{2,4,6,8,t\}$ 
\end{tabular}
\end{table}

These two facts about Fibonacci numbers are documented in Lemma~\ref{fiblemma}.

\begin{lemma}[\cite{stanley2011enumerative,cormen2009}]\label{fiblemma}
Let $k$ be a positive integer. The number of compositions of $k$ into odd parts is precisely $F_k$, and $F_k \ge \phi^{k-2}$.
\end{lemma}

We now state and prove our main theorem.

\begin{theorem}\label{mainthm}
For even $k\ge 2$, $\fs{k}$ is itself non-Hamiltonian but admits $F_k$ many vertex bipartitions $\{S, \bar S\}$ with $s \in S$ for which $S$ and $\bar S$ induce Hamiltonian subgraphs, so the associated 1,2-ATSP instance requires at least $F_k \ge \phi^{n-4}$ SECs.
\end{theorem}

\begin{proof}
First, we show that each composition of $k$ into odd parts gives a vertex bipartition $\{S, \bar S\}$, as shown in Table~\ref{tab:k8}. Let $a_1+a_2+\cdots + a_p$ be a composition of $k$ into odd parts. Since $k$ is even, but each part $a_i$ is odd, the number of parts $p$ must be even. We partition $[k]=\{1,2,\dots, k\}$ into sets $(R_1, R_2,\dots, R_p)$, where the \emph{run} $R_1=\{1,2,\dots, a_1\}$ contains the first $a_1$ vertices, $R_2=\{a_1+1, a_1+2, \dots, a_1+a_2\}$ contains the next $a_2$ vertices, etc. Generally, each run $i$
\begin{align*}
R_i=\{b_i,b_i+1,\dots, f_i\}
\end{align*}
begins at $b_i=\sum_{j=1}^{i-1} a_j+1$ and finishes at $f_i=\sum_{j=1}^{i} a_j$. Observe that the first run $R_1$ begins and finishes at odd numbers, $b_1$ and $f_1$, while $R_2$ begins and finishes at even numbers, $b_2$ and $f_2$, etc. This holds because each $a_i$ is odd. Let
\begin{align*}
S = \{ s \} \cup R_1 \cup R_3 \cup \cdots \cup R_{p-1} \quad\quad
\bar S = \{ t \} \cup R_2 \cup R_4 \cup \cdots \cup R_p.
\end{align*}
We argue that $S$ and $\bar S$ induce Hamiltonian subgraphs, indeed following the precise sequence given above. That is, starting from $s$, traverse run $R_1$ in order, then run $R_3$ in order, and so on, and finally return to $s$. Similarly, for $\bar S$, start from $t$, traverse run $R_2$ in order, then run $R_4$ in order, and so on, and return back to $t$. The edges of these directed cycles exist in $\fs{k}$. Namely, within each run $R_i$, they follow the path edges. When jumping from one run $R_i$ to the next $R_{i+2}$, the finishing vertex $f_i$ and the beginning vertex $b_{i+2}$ have an odd number of vertices between them (namely, $a_{i+1}$), and so $f_i$ and $b_{i+2}$ have the same parity, and so $(f_i, b_{i+2})$ is a parity edge. The first and last edges of $S$'s tour are $(s,1=b_1)$ and $(f_{p-1},s)$, with the former being a path edge, and the latter being a back edge. For $\bar S$'s tour, $(t,b_2)$ is a back edge, and $(f_p=k,t)$ is a path edge. We have thus shown that the $F_k$ many compositions of $k$ into odd parts yield (distinct) suitable partitions of $\fs{k}$'s vertices, and $F_k\ge \phi^{k-2}= \phi^{n-4}$ by Lemma~\ref{fiblemma}.

Finally, we show that $\fs{k}$ is non-Hamiltonian. For contradiction purposes, suppose that it \emph{does} admit a directed Hamiltonian cycle $C$. Denote by $\rho$ the number of path edges in $C$. We count $\rho$ in two ways, finding it to be both odd and even, which will yield our contradiction. Write $C$ as
$(s,h_1,h_2, \dots, h_\alpha, t, q_1, q_2, \dots, q_\beta, s)$.
We have $\alpha, \beta\ge 1$, since there are no edges between $s$ and $t$. The cycle has precisely two back edges, $(t,q_1)$ and $(q_\beta,s)$; all others are path edges or parity edges. Recall that path edges take the form $(i,i+1)$, while parity edges take the form $(i,j)$ with $i<j$ and $j-i$ even. Observe that the path edges cross between the odd numbers (i.e., $1,3,\dots, k-1$) and the even numbers (i.e., $2,4,\dots, k$), while the parity edges and back edges do not. Accordingly, the subpath $P_1=(s,h_1,\dots, h_\alpha,t)$ of $C$, whose interior starts on the odd row and ends on the even row, must have an odd number of path edges, $\rho_1$. Similarly, subpath $P_2=(t,q_1,\dots, q_\beta,s)$ has an odd number of path edges, $\rho_2$. Since $\rho_1$ and $\rho_2$ are odd, their sum $\rho$ is even. 

Next, we show that $\rho$ is odd. 
Classify the vertices $\{1,2,\dots, k\}$ as either type 1 or type 2, depending on whether they lie on subpath $P_1$ or $P_2$. Observe that edges $(s,1)$ and $(k,t)$ are forced into $C$ because these are the only edges exiting $s$ and entering $t$, respectively. Accordingly, these edges belong to subpath $P_1$ and so vertices 1 and $k$ have type 1. Along the path $(s,1,2,\dots, k, t)$ count how many pairs of consecutive vertices have different types. Denote by $c$ this number of ``cut'' edges, each of which points from type 1 to type 2, or from type 2 to type 1. This number $c$ must be even, since we start and end at type 1. Also observe that a given path edge $(i,i+1)$ belongs to $C$ if and only if its vertices have the same type. So, of the $k+1$ path edges, $c$ are cut, and the remaining $\rho=(k+1)-c$ belong to $C$. Since $k+1$ is odd and $c$ is even, their difference $\rho$ is odd.

In conclusion, if $\fs{k}$ admits a Hamiltonian cycle $C$, then the number of path edges $\rho$ in $C$ is odd and even, a contradiction. So, $\fs{k}$ is non-Hamiltonian.
\qed
\end{proof}

By applying a node-splitting reduction, we arrive at Corollary~\ref{cor}. The reduction is similar to that of Jonker and Volgenant~\cite{jonker1983} but exploits properties of $\fs{k}$ to keep the edge costs in $\{1,2,3\}$ and yields a symmetric (indeed \emph{metric}) instance.

\begin{corollary}\label{cor}
For even $k\ge 2$, there exists a metric TSP instance on $n=2(k+2)$ vertices with edge costs in $\{1,2,3\}$ that requires at least $F_k\ge \phi^{n/2-4}$ SECs.
\end{corollary}

\begin{proof}
We apply a node-splitting reduction to $\fs{k}$. Namely, each vertex $v$ of $\fs{k}$ is replaced by two vertices $v$ and $v'$ with an edge between them of cost one. Every directed edge $(u,v)$ in $\fs{k}$ is replaced by undirected edge $\{u',v\}$ of cost two. All other (missing) edges are given cost three. Fig.~\ref{fig:fsu} illustrates this TSP instance $G=(V,E)$, omitting the cost-3 edges for clarity. The only possible obstruction to being metric in a 1,2,3-weighted instance is a 1-1-3 triangle, but no two cost-1 edges share an endpoint, so the instance is metric.

\begin{figure}[ht]
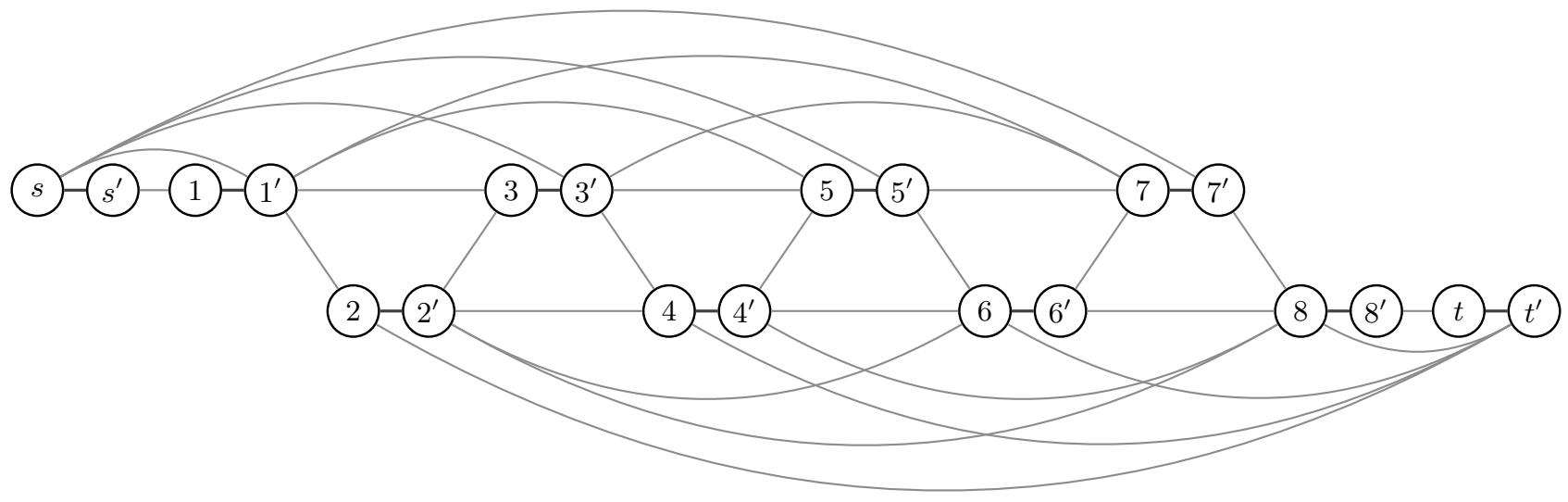

\centering
\resizebox{0.92\columnwidth}{!}
{
\fibgraph[jv, k=8, tikz={fib back/.style={solid}}] 
}
\caption{Applying node-splitting to $\fs{8}$. Edges have three different costs: 1 for edges of the form $\{v,v'\}$, 2 for other drawn edges, and 3 for non-drawn edges. }
\label{fig:fsu}
\end{figure}

Recall that $\fs{k}$ has $F_k$ bipartitions $\{S,\bar S\}$ with both $S$ and $\bar S$ inducing Hamiltonian subgraphs, with the resulting 1,2-ATSP instance $D=(N,A)$ having 
\begin{align}
\tsp{D[S]}+\tsp{D[\bar S]} = |N| < \tsp{D}.\label{tspineq}
\end{align}
These same bipartitions $\{S, \bar S\}$, but in expanded form $\{ T, \bar T\}$, where $T=\{ v,v' \suchthat v \in S\}$ and $\bar T= \{ v, v' \suchthat v \in \bar S\}$, satisfy 
\begin{align*}
\tsp{G[T]}+\tsp{G[\bar T]} = |V|+|N| = 3|N| < \tsp{G}.
\end{align*}
The first equality holds because the directed cycles over $S$ and $\bar S$, when expanded (using the cost-1 edges $\{v,v'\}$), yield cycles over $T$ and $\bar T$ of cost $|V|+|N|=3|N|$. The inequality holds because $\tsp{G}\ge 3|N|$ by the 2-factor relaxation, and if equality held then the presumed tour must use all of the cost-1 edges, and contracting them would give a tour over $D$ of length $|N|$, a contradiction to $\tsp{D}>|N|$ from inequality~\eqref{tspineq}. So, the result follows by Proposition~\ref{cheapbound}.\qed
\end{proof}


Table~\ref{tab:exp-vs-fib} compares the bounds $2^{n/5}$ and $F_k$ on the number of required SECs for the Ball Bearing construction and the symmetric Fibonacci construction for small $n$.
There is little difference at $n=20$. However, at $n=100$, the Ball Bearing construction requires (at least) 1,048,576 SECs while the Fibonacci construction requires (at least) 4,807,526,976. These lower bounds match the values $\R(G)$ reported by the exact algorithm of Vercesi and Buchanan~\cite{vercesi2026} across all tested values up to $n=45$ for the Ball Bearing construction and $n=28$ for the Fibonacci construction, suggesting that our lower bounds are in fact tight.

\begin{table}[ht]
  \caption{Comparing $2^{n/5}$ versus $F_k$ for $n = 2(k+2)$.}
  \centering
  \begin{tabular}{rrrr}
    $n$ & $k$ & $2^{n/5}$ & $F_k$ \\ \hline
    20  & 8  & 16        & 21            \\
    40  & 18 & 256       & 2,584         \\
    60  & 28 & 4,096     & 317,811       \\
    80  & 38 & 65,536    & 39,088,169    \\
    100 & ~48 & ~1,048,576 & ~4,807,526,976 
  \end{tabular}
  \label{tab:exp-vs-fib}
\end{table}

The 1,2,3-TSP instances from Corollary~\ref{cor} have tiny integrality gaps, as noted in Proposition~\ref{prop:fs_ig_sep} and proved in Appendix~\ref{appendix:sep}.

\begin{proposition}\label{prop:fs_ig_sep}
For $k\ge 6$, the 1,2,3-TSP instances $G$ derived from $\fs{k}$ have
\begin{align*}
\frac{\tsp{G}}{\sep{G}} = \frac{\frac{3}{2}n + 1}{\frac{3}{2}n} \to 1~\text{as}~n \to \infty.
\end{align*}
\end{proposition}

\section{Computational Experiments}\label{sec:experiments}

In this section, we assess the empirical hardness of the proposed instances by applying the Concorde TSP solver to them. Experiments were conducted on the fat partition of a shared HPC cluster. Each node has two Intel Xeon E5-2650 v3 CPUs (2.30 GHz, 10 cores each, 20 cores total) and 128 GB of RAM. Jobs were limited to 48 hours. The code was written in Python, using PyConcorde~\cite{pyconcorde} (and, as such, QSOpt as an optimizer). The code (including generators for the planar embeddings) and results are available at \href{https://anonymous.4open.science/r/DFJ_requires_exponentially_many_SECs-3386/}{this link}.

\begin{figure}[ht]
\centering
    \begin{subfigure}[b]{0.48\textwidth}
    \centering
    \includegraphics[width=\textwidth]{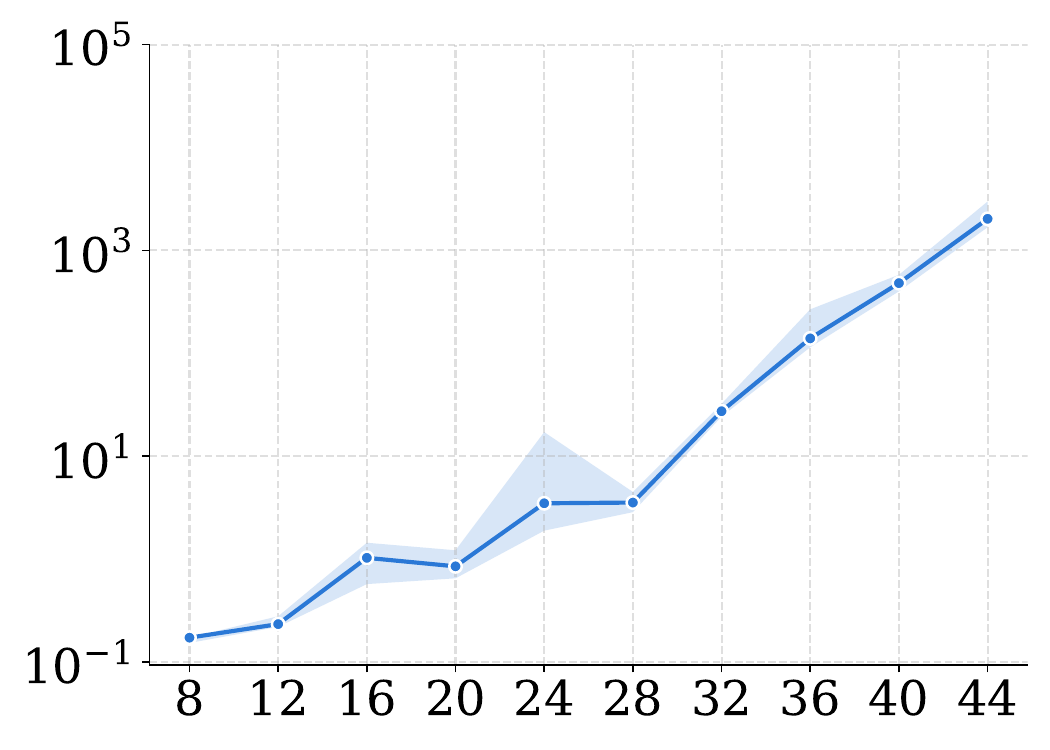}
    \caption{Fibonacci instances, 1,2,3-weighting. \label{fig:fibonacci}}
    \end{subfigure}
\quad
    \begin{subfigure}[b]{0.48\textwidth}
    \centering
    \includegraphics[width=\textwidth]{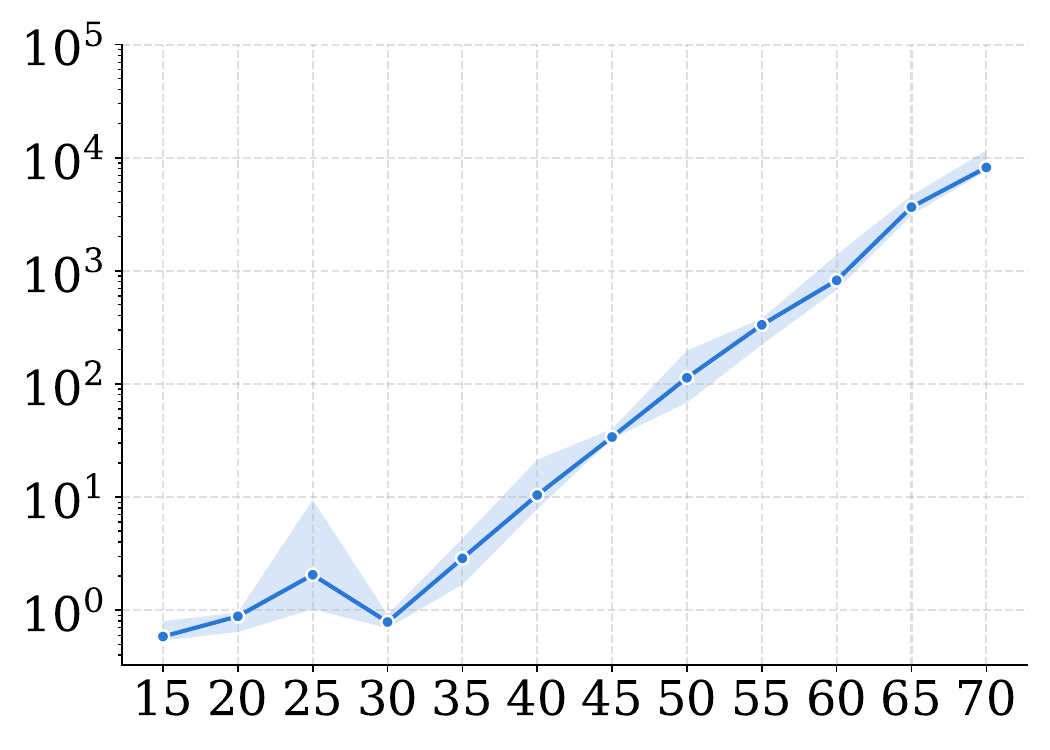}
    \caption{BB instances, 1,2-weighting.\label{fig:bb}}
    \end{subfigure}
\quad
    \begin{subfigure}[b]{0.48\textwidth}
    \centering
    \includegraphics[width=\textwidth]{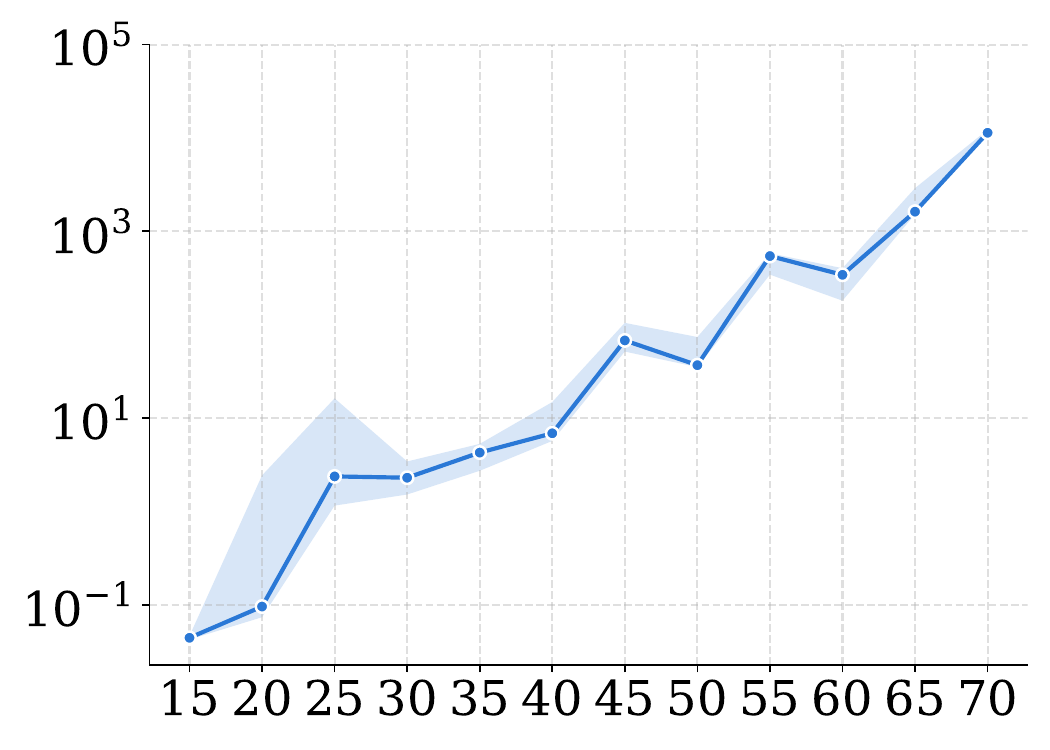}
    \caption{BB instances, Euclidean metric.\label{fig:bb_eucl}}
    \end{subfigure}
\quad
\begin{subfigure}[b]{0.48\textwidth}
    \centering
    \includegraphics[width=\textwidth]{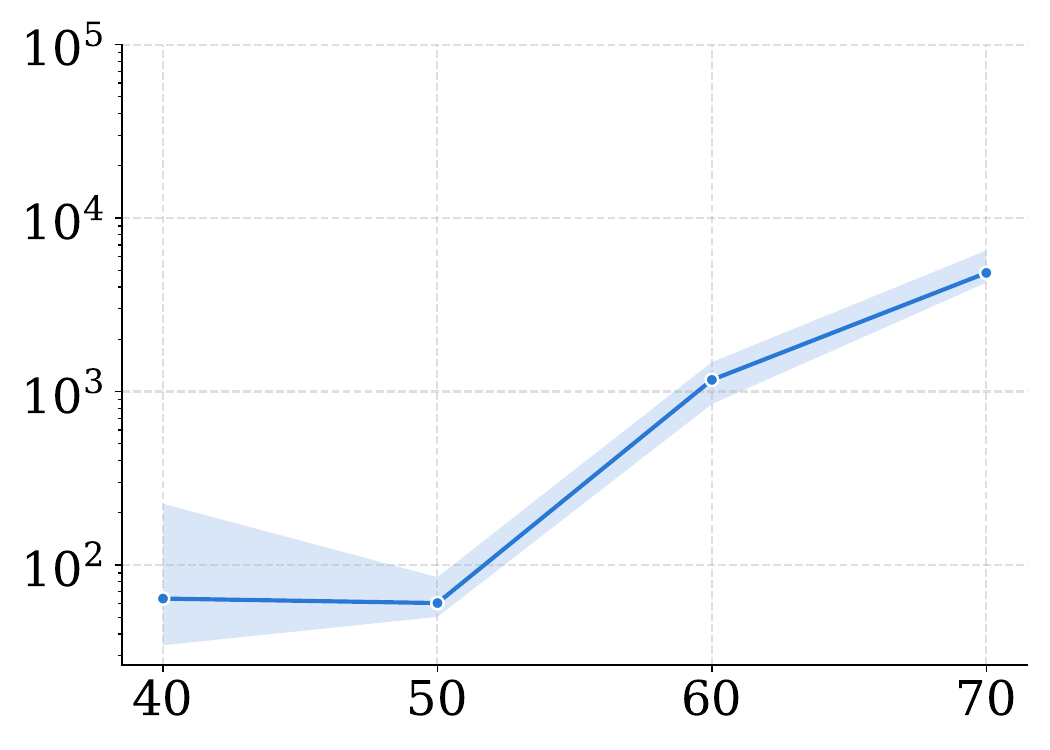}
    \caption{BB instances, Maximum metric.\label{fig:bb_max}}
    \end{subfigure}
\quad
\begin{subfigure}[b]{0.48\textwidth}
\centering
\includegraphics[width=\textwidth]{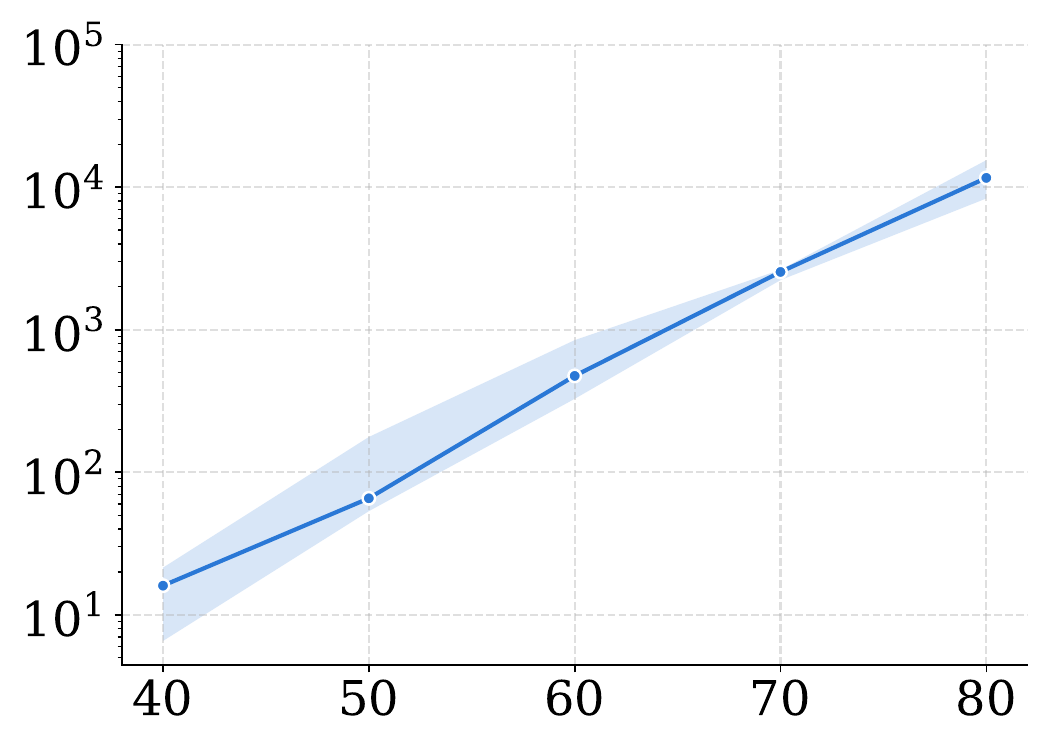}
\caption{BB instances, Rectilinear metric.\label{fig:bb_rect}}
\end{subfigure}
\caption{Behavior of Concorde: number of vertices versus time in sec ($\log_{10}$ scale).}\label{fig:trends}
\end{figure}

We consider five classes of TSP instances, including four Ball Bearing constructions (1,2-TSP, Euclidean, Maximum, Rectilinear) and one Fibonacci construction (1,2,3-TSP)
for different values of $n$. We apply Concorde to each instance using 10 different seeds and plot the median solve time and interquartile range in Fig.~\ref{fig:trends}. For the Fibonacci construction, the job was killed when $k>20$ ($n>44$) due to a memory issue; all other runs either solved to optimality or exceeded the time limit. 

Running times are consistent with many other hard instances of similar size in the metric context (e.g., HardTSPLIB~\cite{vercesi2023generation}). 
In comparison to the tetrahedron instances of Hougardy and Zhong \cite{hougardy2021hard}, the Euclidean embedding of the Ball Bearing construction (scaled by 1000) gives results that are two orders of magnitude slower (for 70 nodes >10,000s here versus $\le$ 100s in Fig.~12 and 13 of~\cite{hougardy2021hard}), leading to the ``hardest'' Euclidean instances currently available in the literature. Fig.~\ref{fig:bb_max} and \ref{fig:bb_rect} give runtimes for the Maximum and Rectilinear embeddings. 
The Rectilinear times are faster than those reported by Zhong~\cite{zhong2025} but are still much slower than those of TSPLIB instances. 

We remark that the proposed instances are not hard in the usual computational complexity sense. As we saw, the 1,2-weighting of the Ball Bearing construction has $\tsp{G}=n+1$, and a matching lower bound is achieved by exploiting the fact that $\bb{k}$ is non-Hamiltonian and adding the cut $x(E(\bb{k}))\le n-1$ to the LP relaxation over the degree-two constraints and 0,1-bounds. This specialized ``hack'' would speed up Concorde but is not algorithmically satisfying.

\section{Conclusion and Future Work}\label{sec:conclusion}

This paper constructs pathological TSP instances that require exponentially many subtour elimination constraints, with growth rates of at least $2^{n/5}$ for instances in the plane under the Euclidean, Maximum, and Rectilinear metrics, and growth rates of $\phi^n$ for 1,2-ATSP and $\phi^{n/2}$ for metric TSP. 

Future work may consider questions like: What is the expected number of required SECs for randomly-generated TSP instances (e.g., in a unit square) or for ``typical'' instances (however defined)? Is the growth polynomial or exponential? Is the $\phi^n$ growth in SECs for ATSP optimal, or do there exist instances with faster growth rate? Can any nontrivial upper bound, say, $c^n$ for $c<2$, be proven? 

\clearpage

\paragraph{Acknowledgments.} 
The research of E.~Vercesi was supported by the Swiss National Science Foundation [Grant 200021\_212929/1] ``Computational methods for integrality gaps analysis.'' The research of
A.~Buchanan is based on work supported the Air Force Office of Scientific Research [Grant FA9550-25-1-0277]. 
The authors developed and refined the constructions in concert with Claude (e.g., Opus 5.5). Claude was also used to generate TikZ figures. {For the experiments, GitHub Copilot was used for code completion and for writing the documentation.} {However, all text in this paper, including prose, theorem statements, and proofs, was written solely by the (human) authors.}

\bibliographystyle{plain}
\bibliography{tsp}


\clearpage
\newpage

\appendix

\section*{Appendix}

\section{Integrality Gap Proofs}\label{appendix:sep}

\begin{proof}[of Proposition~\ref{prop:bb_ig_sep}]
We know $\tsp{G} \ge n+1$ by the proof of Theorem~\ref{bb12}. For $\tsp{G}\le n+1$, consider a gadget $g=\{i,b,p_1,p_2,p_3\}$ and a tour constructed by following the outer ring clockwise ($p_2, p_3, p'_1, \dots, p_1$), crossing into the inner ring at ball vertex $b$, and then following the inner ring starting from $i'$, picking up all other ball vertices along the way until reaching back to $i$, and closing the tour at $p_2$. All tour edges exist in $\bb{k}$ except final edge $ip_2$, giving cost $n+1$. 

We prove $\sep{G}=n$. The degree-two constraints~\eqref{deg2} imply $\sum_{e \in E} x_e = n$, and 1,2-weighting gives cost coefficients $c_e \ge 1$, so all $x\ge 0$ satisfy
\begin{align*}
\sum_{e \in E} c_e x_e \ge \sum_{e \in E} x_e = n.
\end{align*}
In particular, this holds for any $x$ in the subtour LP.
To prove $\sep{G} \le n$, consider the 2-factors from Fig.~\ref{two2f}. The 2-factor $F$ puts the ball vertices on the inner cycle, while $F'$ puts them on the outer cycle. Their midpoint $x^*=\frac{1}{2}(x^{F}+x^{F'})$ is half-integral and by convexity has the same cost $n$ and satisfies the degree-two constraints and 0,1-bounds. 

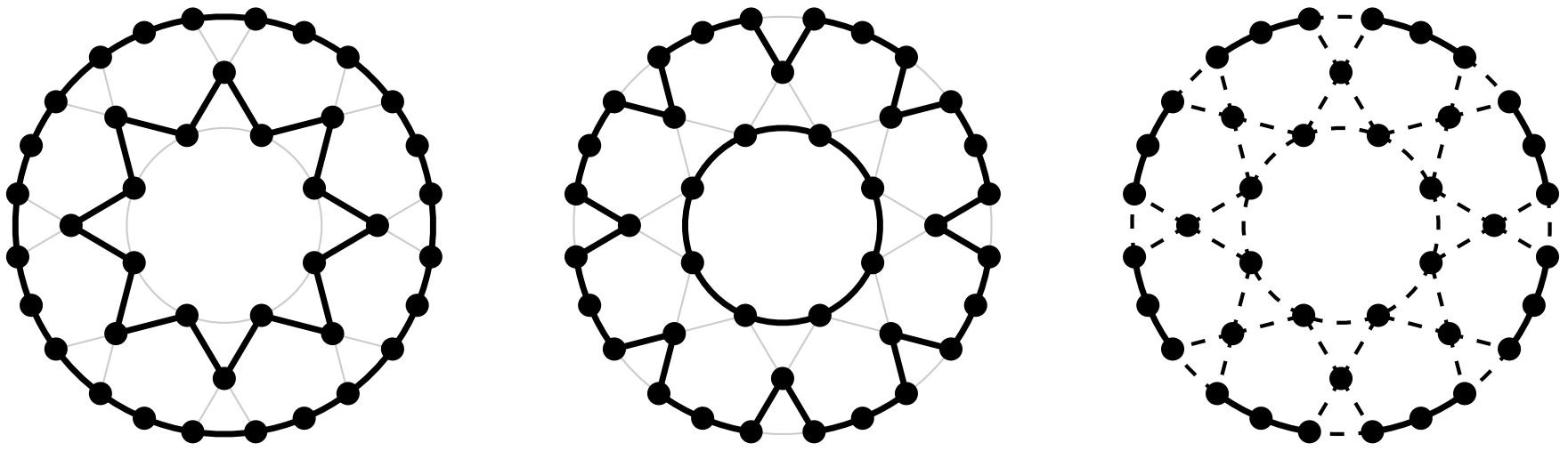
\begin{figure}[ht]
  \centering
  \begin{subfigure}{0.32\linewidth}\centering
    \begin{tikzpicture}
      \TwoNFiveRing[gadgets=8, highlight=-1, light=black, labels=false, edge mode=F1, width=\linewidth]
    \end{tikzpicture}
    \caption{$F$: balls on inner cycle.}
  \end{subfigure}\hfill
  \begin{subfigure}{0.32\linewidth}\centering
    \begin{tikzpicture}
      \TwoNFiveRing[gadgets=8, highlight=-1, light=black, labels=false, edge mode=F2, width=\linewidth]
    \end{tikzpicture}
    \caption{$F'$: balls on outer cycle.}
  \end{subfigure}\hfill
  \begin{subfigure}{0.32\linewidth}\centering
    \begin{tikzpicture}
      \TwoNFiveRing[gadgets=8, highlight=-1, light=black, labels=false, edge mode=lp, width=\linewidth]
    \end{tikzpicture}
    \caption{their midpoint $x^*$.}
  \end{subfigure}
  \caption{Subtour LP solution as midpoint of 2-factors (Ball Bearing).}
  \label{two2f}
\end{figure}

It remains to show that $x^*$ satisfies an arbitrary SEC $x(\delta(S))\ge 2$. In the first case, suppose that $S$ coincides with the vertex set of a cycle of $F$ or a cycle of $F'$. In this case, the \emph{other} 2-factor crosses the cut $(S, \bar S)$ a total of $2k$ times (twice for each ball), in which case 
\begin{align*}
x^*(\delta(S))=\frac{1}{2} \left( x^{F}(\delta(S))+x^{F'}(\delta(S)) \right )=\frac{1}{2}(2k)=k\ge 3 > 2.
\end{align*}
Otherwise, if $S$ does \emph{not} coincide with a cycle of $F$ nor a cycle of $F'$, then $F$ and $F'$ each cross the cut $(S, \bar S)$ at least twice, so
\begin{align*}
x^*(\delta(S))=\frac{1}{2} x^{F}(\delta(S))+\frac{1}{2}x^{F'}(\delta(S))\ge 1+1=2. 
\end{align*}
Thus, the midpoint solution $x^*$ proves $\sep{G}\le n$. \qed
\end{proof}

\begin{proof}[of Proposition~\ref{prop:fs_ig_sep}]
As before, let $D=(N,A)$ be the 1,2-ATSP instance and $G=(V,E)$ be the 1,2,3-TSP instance with $n=|V|=2|N|$ vertices. By the proof of Corollary~\ref{cor} we have $\tsp{G} > \frac{3}{2}n$. The integrality of the cost coefficients $c_e$ and even parity of $n$ then imply $\tsp{G}\ge \frac{3}{2}n+1$. To see $\tsp{G}\le \frac{3}{2}n+1$, consider the straightforward tour $(s,s',1,1', 2, 2',\dots, k, k', t, t', s)$, which uses $n/2$ edges of cost 1, $n/2-1$ edges of cost 2, and one final edge $\{t',s\}$ of cost 3, giving a total cost of $1(n/2) + 2(n/2-1) + 3(1)=\frac{3}{2}n+1$, as desired. 

We prove $\sep{G}=\frac{3}{2}n$. For $\sep{G}\ge \frac{3}{2}n$, see that each vertex $v$ has only one incident edge of cost 1; all others have cost 2 or 3. So, the degree-two constraints~\eqref{deg2} and 0,1-bounds force a local cost of $\sum_{e \in \delta(v)} c_e x_e \ge 2+1=3$ and total cost of
\begin{align*}
\sum_{e \in E} c_e x_e = \frac{1}{2} \sum_{v \in V} \sum_{e \in \delta(v)} c_e x_e \ge \frac{1}{2} \sum_{v \in V} 3 = \frac{3}{2}n.
\end{align*}
In particular, this holds for points in the subtour LP. To prove $\sep{G}\le \frac{3}{2}n$, consider the 2-factors from Fig.~\ref{fig:sep-midpoint}. The 2-factor $F$ is based on the all-ones composition $1+1+\cdots +1$, while $F'$ uses $3+3+1+\cdots +1$. (Hence the $k\ge 6$ assumption.) Their midpoint $x^*=\frac{1}{2}( x^{F}+x^{F'})$ is half-integral and by convexity has the same cost $\frac{3}{2}n$ and satisfies the degree-two constraints and 0,1-bounds. 

\colorlet{FA}{blue!70!black}     
\colorlet{FAt}{blue!45}          
\colorlet{FB}{orange!85!black}   
\colorlet{FBt}{orange!55}        
\tikzset{
  fib edge/.style={draw=black!18,line width=0.5pt},    
  fib gadget/.style={draw=black!30,line width=0.7pt},  
  half/.style={line width=1.2pt,dash pattern=on 3.5pt off 2pt},
}
\fibsetup{jv, k=6, fill by side=false}
 
 
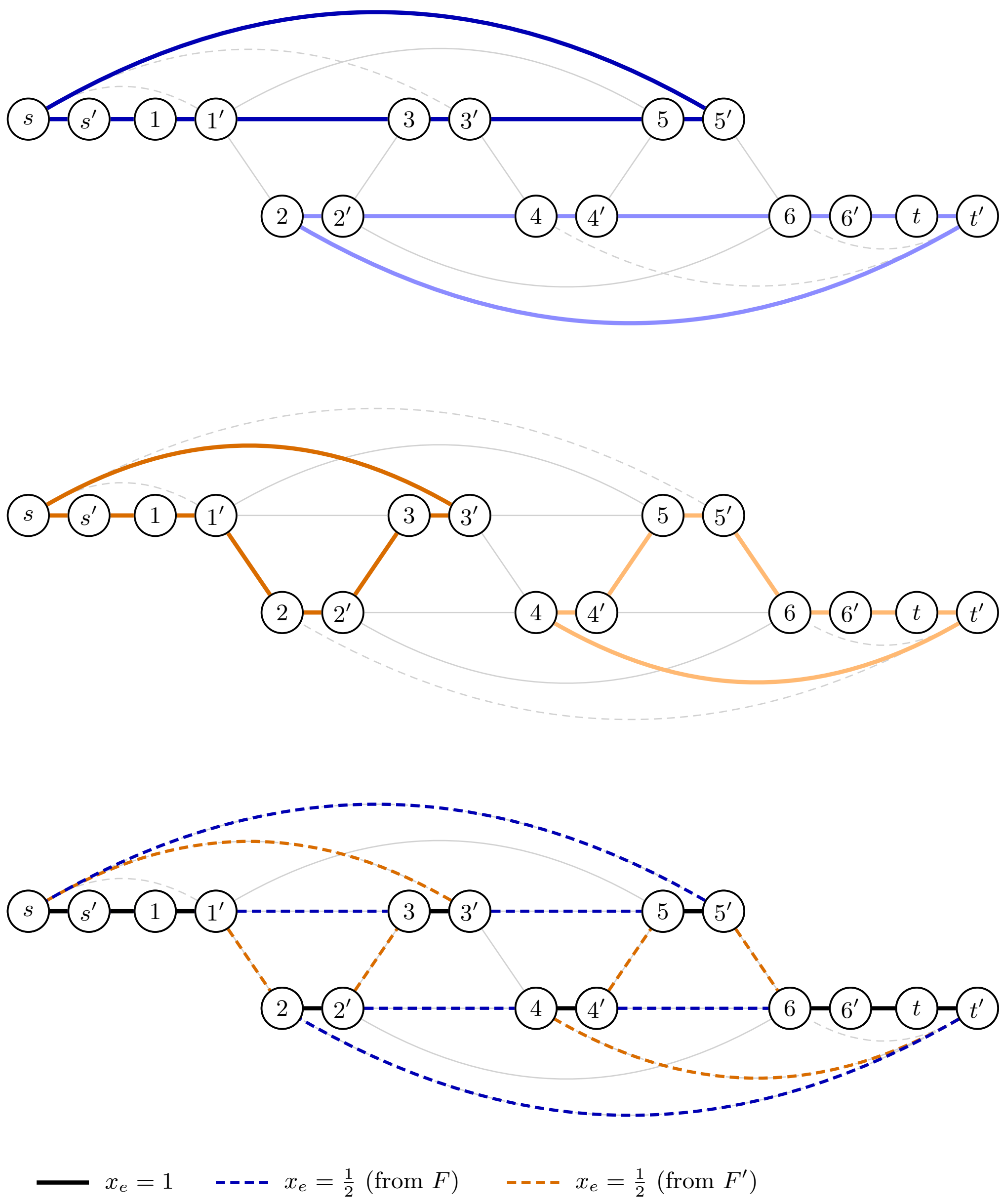
\begin{figure}[ht]
\centering
\begin{subfigure}{\linewidth}\centering
  \tikzset{fib a cycle/.style={draw=FA,line width=1.6pt},
           fib b cycle/.style={draw=FAt,line width=1.6pt}}
  \fibgraph[composition={1,1,1,1,1,1}]

\end{subfigure}
 
\medskip
\begin{subfigure}{\linewidth}\centering
  \tikzset{fib a cycle/.style={draw=FB,line width=1.6pt},
           fib b cycle/.style={draw=FBt,line width=1.6pt}}
  \fibgraph[composition={3,3}]

\end{subfigure}
 
\medskip
\begin{subfigure}{\linewidth}\centering
  \begin{tikzpicture}
    \tikzset{fib a cycle/.style={draw=FB,half}, fib b cycle/.style={draw=FB,half}}
    \fibgraphdraw[composition={3,3}]
    \begin{scope}[fib a side/.style={fill=white}, fib b side/.style={fill=white}]
      \tikzset{fib a cycle/.style={draw=FA,half}, fib b cycle/.style={draw=FA,half}}
      \fibgraphdraw[composition={1,1,1,1,1,1}, only cycles, fill by side]
    \end{scope}
    \foreach \v in {s,1,2,3,4,5,6,t} \draw[black,line width=1.6pt] (fib-\v) -- (fib-\v p);
    \draw[black,line width=1.6pt] (fib-sp) -- (fib-1);
    \draw[black,line width=1.6pt] (fib-6p) -- (fib-t);
    \begin{scope}[shift={($(current bounding box.south west)+(0.4,-0.45)$)},font=\footnotesize]
      \draw[black,line width=1.6pt] (0,0) -- (0.7,0); \node[anchor=west] at (0.8,0) {$x_e=1$};
      \draw[FA,half] (2.4,0) -- (3.1,0); \node[anchor=west] at (3.2,0) {$x_e=\tfrac12$ (from $F$)};
      \draw[FB,half] (6.3,0) -- (7.0,0); \node[anchor=west] at (7.1,0) {$x_e=\tfrac12$ (from $F'$)};
    \end{scope}
  \end{tikzpicture}
\end{subfigure}
\caption{Subtour LP solution as midpoint of 2-factors (Fibonacci).}
\label{fig:sep-midpoint}
\end{figure}

It remains to show that $x^*$ satisfies an arbitrary SEC $x(\delta(S)) \ge 2$. Observe that $F$ has two cycles: one over the odd row 
\begin{align*}
V_1\coloneqq\{s, s', 1, 1', 3, 3', \dots, (k-1), (k-1)' \},
\end{align*}
and one over the even row 
\begin{align*}
V_2\coloneqq\{ 2, 2', 4, 4', \dots, k, k', t, t' \}. 
\end{align*}
Meanwhile, $F'$ is similar, with the only changes being that $2$ and $2'$ are moved to the odd row, and that $5$ and $5'$ are moved to the even row. Denote by $V'_1$ and $V'_2$ the associated vertex sets.
If $S \notin \{V_1, V_2, V'_1, V'_2\}$, then each of $F$ and $F'$ crosses the cut $(S, \bar S)$ at least twice, so
\begin{align*}
x^*(\delta(S))=\frac{1}{2} x^{F}(\delta(S))+\frac{1}{2}x^{F'}(\delta(S))\ge 1+1=2. 
\end{align*}
Otherwise, we have $S\in \{V_1, V_2, V'_1, V'_2\}$. Now, if $S\in \{V_1, V_2\}$, then $F'$ crosses the cut four times, specifically at edges $\{1',2\}, \{2',3\},\{4',5\},\{5',6\}$, so
\begin{align*}
x^*(\delta(S))=\frac{1}{2} \left( x^{F}(\delta(S))+x^{F'}(\delta(S)) \right )=\frac{1}{2}(4)=2.
\end{align*}
The case $S\in\{V'_1,V'_2\}$ is analogous, at edges $\{t',2\},\{2',4\},\{3',5\}$ and lastly $\{5',s\}$ if $k=6$ (or $\{5',7\}$ if $k\ge 8$). Thus, $\sep{G} \le \frac{3}{2}n$ by the midpoint $x^*$. \qed
\end{proof}

\clearpage
\newpage

\section{Extra Figures}\label{appendix:figs}

\begin{figure}
\centering 
\begin{tikzpicture}
\StackRing[a=1, b=1, small, grid, width=\linewidth]
\end{tikzpicture}
\caption{A square coin embedding of $\bb{8}$ for Maximum metric.}
\label{fig:simplestmax8}
\end{figure}
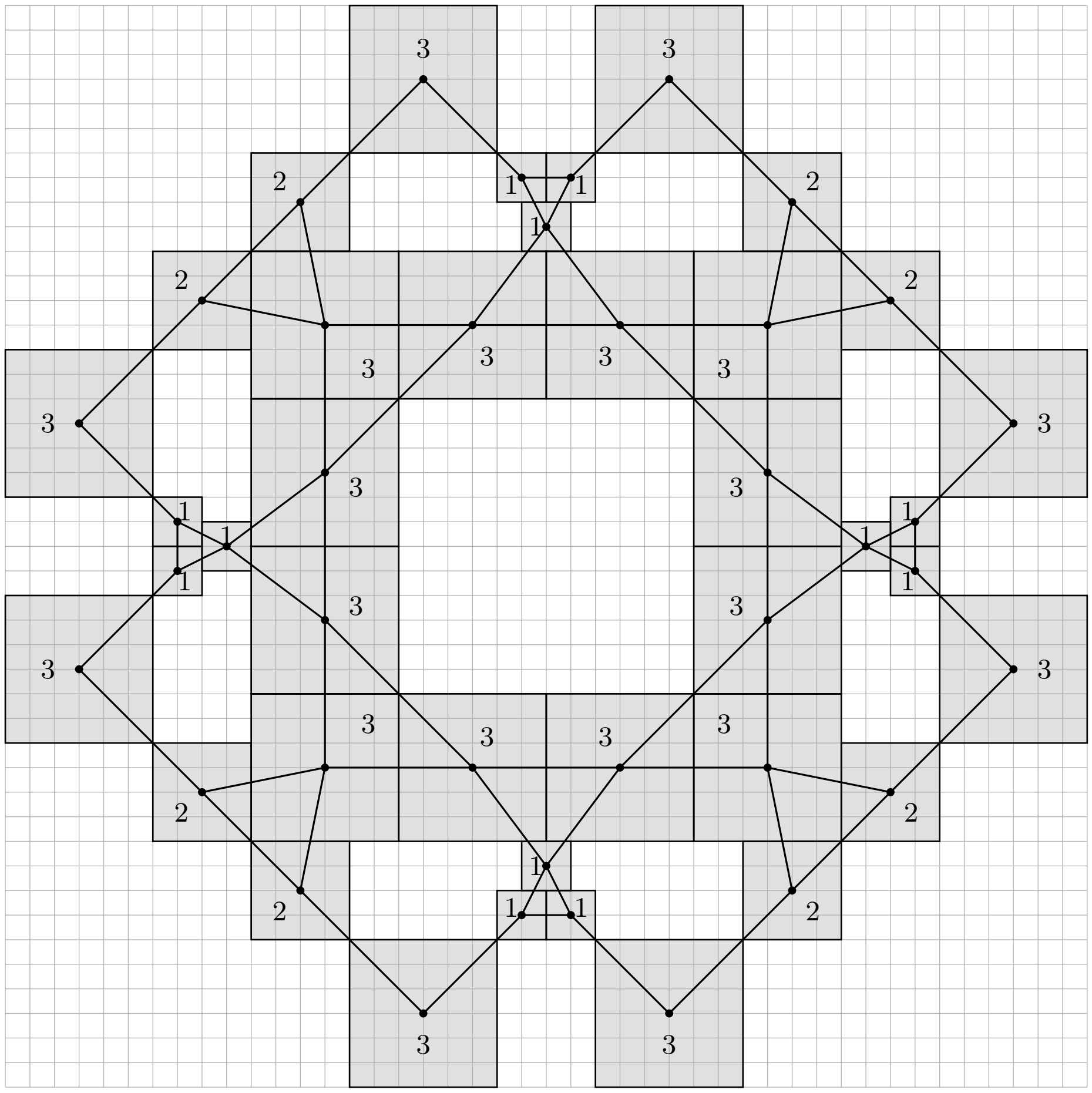

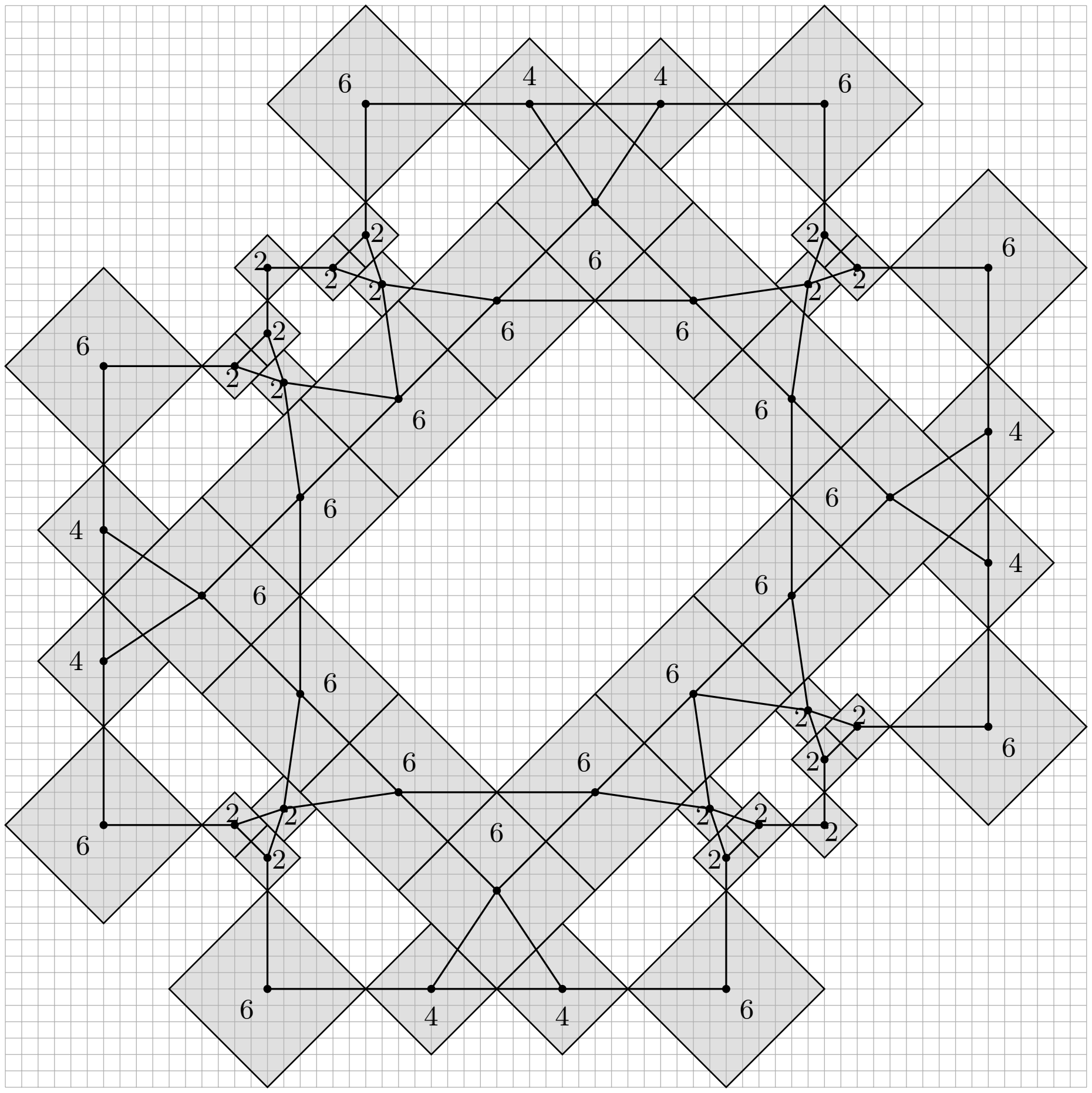
\begin{figure}
\centering 
\begin{tikzpicture}
\StackRing[a=2, b=1, small, width=\linewidth, rectilinear integer, grid]
\end{tikzpicture}
\caption{A diamond coin embedding of $\bb{10}$ for Rectilinear metric. Integer coordinates are preserved via doubling.}
\label{fig:simplestrect}
\end{figure}

\end{document}